\documentclass[prb, superscriptaddress, twocolumn]{revtex4}

\usepackage{amsmath, amssymb, braket}
\usepackage{mathrsfs}
\usepackage{epstopdf}
\usepackage{graphicx, pdflscape, array, makecell, multirow}
\usepackage[sort&compress]{natbib}
\usepackage{natmove}
\usepackage[pdftex = true, colorlinks = true, citecolor = blue, bookmarks = true]{hyperref}
\usepackage{hypernat}
\usepackage[utf8]{inputenc}
\usepackage[OT4]{fontenc}

\newcommand{\asop}{\ensuremath{\hat{\mathscr{A}}}}

\newcommand{\beq}{\begin{equation}}

\newcommand{\eeq}{\end{equation}}

\newcommand{\bec}{\begin{center}}

\newcommand{\eec}{\end{center}}

\begin{document}


\title{Derivation of the Supermolecular Interaction Energy from the Monomer Densities \\ in the Density Functional Theory}
\date{\today}
\author{Łukasz Rajchel}
\email{lrajchel@tiger.chem.uw.edu.pl}
\affiliation{Department of Chemistry, Oakland University, Rochester, Michigan 48309-4477, USA}
\affiliation{Faculty of Chemistry, University of Warsaw, 02-093 Warszawa, Pasteura 1, Poland}
\author{Piotr S. Żuchowski}
\affiliation{Department of Chemistry, Durham University, South Road, Durham DH1 3LE, United Kingdom}
\author{Małgorzata M. Szczęśniak}
\affiliation{Department of Chemistry, Oakland University, Rochester, Michigan 48309-4477, USA}
\author{Grzegorz Chałasiński}
\email{chalbie@tiger.chem.uw.edu.pl}
\affiliation{Department of Chemistry, Oakland University, Rochester, Michigan 48309-4477, USA}
\affiliation{Faculty of Chemistry, University of Warsaw, 02-093 Warszawa, Pasteura 1, Poland}



\begin{abstract}

The density functional theory (DFT) interaction energy of a dimer is rigorously derived from the monomer densities. To this end, the supermolecular energy bifunctional is formulated in terms of mutually orthogonal sets of orbitals of the constituent monomers. The orthogonality condition is preserved in the solution of the Kohn-Sham equations through the Pauli blockade method. Numerical implementation of the method provides interaction energies which agree with those obtained from standard supermolecular calculations within less than 0.1~\% error for three example functionals: Slater-Dirac, PBE0 and B3LYP, and for two model van der Waals dimers: Ne$_2$ and (C$_2$H$_4$)$_2$, and two model H-bond complexes: (HF)$_2$ and (NH$_3$)$_2$.

\end{abstract}

\maketitle


\section{Introduction}

Deriving the dimer interaction energy via mutual polarization of constituent monomers is important both from the fundamental perspective and from a practical point of view. In particular, it may aid the understanding how the non-covalent systems are described in the density functional theory which is one of the most problematic issues of the electronic structure theory. The major problem of DFT as applied to the van der Waals systems is a wrong description of dispersion forces~\cite{angyan_gerber_savin_toul:2005, dion_rydb_schr_langr_lund:2004, dobson_wang_dinte_mclenna_le:2005}. Surprisingly enough, little has been done to better understand the performance of supermolecular interaction energy in the framework of DFT. For the Hartree-Fock (HF) interaction energy, such an approach has been pioneered by \citet{mor:1971} in the 1970s, and a decade later, inspired by the work of~\citet{sadlej_hhf:1980}, rigorously derived by~\citet{gut_pauli:1988} (see also Ref.~\cite{olsz_gut_piela:1990}). The perturbation approach within the symmetry-adapted perturbation theory~(SAPT) formalism was also extensively exploited in this context~\cite{jez_mosz_szal:1994, jeziorska_jez_ciz:1987}. In the age of DFT, it is highly desirable to develop such an approach also for the density functional formalism. Approximate DFT treatments have already been advanced by \citet{cortona:1991} and \citet{wesol_warchel:1993}; see also recent energy decomposition schemes proposed in Refs.~\citet{cyb_seversen:2003, reinh_piqueman_savin:2008, su_li:2009}, and Refs. therein, as well as the density functional formulation of SAPT~\cite{mis_jez_szal:2003, hess_jansen:2003}.

The goal of this work is to derive rigorously the supermolecular density functional theory~(DFT) interaction energy via the mutual polarization of the monomer densities. To this end the supermolecular (dimer) energy functional is expressed in terms of mutually orthogonalized sets of the Kohn-Sham~(KS) orbitals of the constituent monomers. The coupled KS equations are next solved iteratively, by using the Pauli blockade~(PB) technique of~\citet{gut_pauli:1988}. The correctness of the derivation is demonstrated by comparing the interaction energy calculated from the equations developed here and the supermolecular interaction energies. The DFT approximation to the Heitler London interaction energy, based on the decomposition of the interaction energy introduced in this paper, is also discussed.

\section{Theory}
\label{sec:theory}

\newcommand{\kinop}{\ensuremath{-\frac{1}{2} \Delta_\mathbf{r}}}
\newcommand{\pkinop}{\ensuremath{\vphantom{-\frac{1}{2} \Delta_\mathbf{r}}}}

In this paper we consider the interaction between two closed-shell systems, however, the generalization for high-spin open-shell systems and clusters is straightforward. The supermolecular interaction energy in terms of DFT can be defined as the difference between the total energies of the dimer AB and the individual monomers~A and~B, separated to infinity:
	\beq
		E_\mathrm{int}^\mathrm{DFT} = E_\mathrm{AB}^\mathrm{DFT} - E_\mathrm{A}^\mathrm{DFT} - E_\mathrm{B}^\mathrm{DFT} 
		\label{eq:Eint_DFT}.
	\eeq
It was demonstrated by~\citet{gut_pauli:1988} that the HF supermolecular interaction energy may be exactly recovered  by solving the HF equations for monomers in the presence of the external perturbation, consisting of the electrostatic potential and the non-local exchange potential generated by the second monomer. They have also proposed a convenient computational scheme for the PB method in terms of mutually orthogonalized A~and~B orbitals with the penalty operator forcing the orthogonality of  monomers' occupied orbitals. In this section we derive an analogous formalism in terms of appropriately modified KS equations and monomer densities.

We begin with KS equations for the isolated monomers which yield the starting orbitals~$\left\{ a_i^0 \right\}_{i \in A}$ and~$\left\{ b_k^0 \right\}_{k \in B}$. The orbitals of monomer~A are the solutions of the following eigen equation:
	\beq
		\hat{f}_\mathrm{A}^{\mathrm{KS},0}(\mathbf{r}) a_i^0(\mathbf{r}) = \epsilon_{\mathrm{A}, i}^0 a_i^0(\mathbf{r}),
		\label{eq:fKS0}
	\eeq
and the analogous equations hold for the monomer~B. The KS~operator of~Eq.~\eqref{eq:fKS0} is written as
	\beq
		\hat{f}_\mathrm{A}^{\mathrm{KS},0}(\mathbf{r}) = -\frac{1}{2} \Delta_\mathbf{r} + v_\mathrm{A}^\mathrm{ne}(\mathbf{r}) + \hat{\jmath}_\mathrm{A}(\mathbf{r}) + v_\mathrm{A}^\mathrm{xc}(\mathbf{r}),
	\eeq
where the monomer~A nuclear potential is
	\beq
		v_\mathrm{A}^\mathrm{ne}(\mathbf{r}) = - \sum_{\alpha = 1}^{\mathscr{N}_\mathrm{A}} \frac{Z_\alpha}{|\mathbf{r} - \mathbf{R}_\alpha|}
	\eeq
with~$\mathscr{N}_\mathrm{A}$ being the number of monomer~A nuclei, each described by its position~$\mathbf{R}_\alpha$ and charge~$Z_\alpha$. Its coulombic potential reads
	\beq
		\hat{\jmath}_\mathrm{A}(\mathbf{r}) = \int_{\mathbb{R}^3} \frac{\rho_\mathrm{A}^0(\mathbf{r}')}{|\mathbf{r} - \mathbf{r}'|} \, d^3 \mathbf{r}'.
	\eeq
The total energy of monomer A can be written as
	\beq
		\begin{split}
			E_\mathrm{A} \left[ \rho_\mathrm{A}^0 \right] & = T^\mathrm{s} \left[ \rho_\mathrm{A}^0 \right] + V_\mathrm{A}^\mathrm{ne} \left[ \rho_\mathrm{A}^0 \right] +
				J \left[ \rho_\mathrm{A}^0 \right] + \\
			 & + E^\mathrm{xc} \left[ \rho_\mathrm{A}^0 \right] + V_\mathrm{A}^\mathrm{nn}. \label{eq:PB_fundef_TsVneJExc}
		\end{split}
	\eeq
The functional~\eqref{eq:PB_fundef_TsVneJExc} comprises the non-interacting kinetic energy:
	\beq
		T^\mathrm{s} \left[ \rho_\mathrm{A}^0 \right] = 2 \sum_{i \in A} \Bra{ a_i^0 \pkinop } \kinop \Ket{ \pkinop a_i^0 },
		\label{eq:PB_Ts_fun}
	\eeq
with $A$ being the set of the indices of occupied orbitals of the monomer~A, nuclear-electron attraction energy:
	\beq
		V_\mathrm{A}^\mathrm{ne} \left[ \rho_\mathrm{A}^0 \right] = \int_{\mathbb{R}^3} v_\mathrm{A}^\mathrm{ne}(\mathbf{r}) \rho_\mathrm{A}^0(\mathbf{r}) \, d^3 \mathbf{r},
	\eeq
coulombic energy
	\beq
		J \left[ \rho_\mathrm{A}^0 \right] =
			\frac{1}{2} \int_{\mathbb{R}^3} \int_{\mathbb{R}^3} \frac{\rho_\mathrm{A}^0(\mathbf{r}_1) \rho_\mathrm{A}^0(\mathbf{r}_2)}{r_{12}} \, d^3 \mathbf{r}_1 d^3 \mathbf{r}_2,
	\eeq
exchange-correlation~(xc) energy
	\beq
		E^\mathrm{xc} \left[ \rho_\mathrm{A}^0 \right] =
			\int_{\mathbb{R}^3} F^\mathrm{xc} \Big( \rho_\mathrm{A}^0(\mathbf{r}); \Big\{ \nabla_\mathbf{r} \rho_\mathrm{A}^0(\mathbf{r}); \ldots \Big\} \Big) \, d^3\mathbf{r}
	\eeq
which is evaluated through the numerical integration of the~$F^\mathrm{xc}$ integrand on a~grid of points around monomer~A, and the nuclear-nuclear repulsion term
	\beq
		V_\mathrm{A}^\mathrm{nn} = \sum_{\alpha = 1}^{\mathscr{N}_\mathrm{A} - 1} \sum_{\beta = \alpha + 1}^{\mathscr{N}_\mathrm{A}} \frac{Z_\alpha Z_\beta}{R_{\alpha \beta}}
	\eeq
which is constant for a~fixed geometry. The density of monomer~A is
	\beq
		\rho_\mathrm{A}^0(\mathbf{r}) = 2 \sum_{i \in A} \left| a_i^0(\mathbf{r}) \right|^2.
	\eeq
Similar expressions can be written for monomer~B.

The original, isolated-monomer orbital sets~$\left\{ a_i^0 \right\}_{i \in A}$ and~$\left\{ b_k^0 \right\}_{k \in B}$ are not mutually orthogonal. To proceed, it is also important to introduce the set of orthonormalized orbitals which are obtained by using Löwdin symmetric orthonormalization~\cite{lowin:1950}. The quantities expressed in the orthonormalized orbitals are henceforth marked with tilde. One should remember that the orthonormalization leaves the total density of the dimer unchanged. However, it does change the monomer densities into the densities deformed by the presence of the interacting partner.

In the PB method the zeroth-order wavefunction of the dimer is the wavefunction of the system in the absence of molecular interaction. It is constructed from the antisymmetrized product of the orthogonalized occupied orbitals of the monomers~A and~B. In case of KS equations for dimer the KS determinant can be constructed in the same manner as:
	\beq
		\psi_\mathrm{AB}^0 = \asop \tilde{\psi}_\mathrm{A}^0 \tilde{\psi}_\mathrm{B}^0,
		\label{eq:PB_psiAB_NA_A_B}
	\eeq
where~$\tilde{\psi}_\mathrm{A}^0$ and~$\tilde{\psi}_\mathrm{B}^0$ are KS determinants of monomers~A and~B, respectively. Since the determinants are constructed from orthonormalized orbitals, the $ \psi_\mathrm{AB}^0$ is normalized.

It can be easily shown that the zeroth-order density of the system can be simply written as a sum of monomer densities expressed in terms of orthonormalized orbitals,
	\beq
		\rho_\mathrm{AB}^0 = \tilde{\rho}_\mathrm{AB}^0 = \tilde{\rho}_\mathrm{A}^0 + \tilde{\rho}_\mathrm{B}^0.
		\label{eq:PB_trhoAB_add}
	\eeq
Note that~\eqref{eq:PB_trhoAB_add} does not hold for the densities obtained from nonorthonormal orbitals of the dimer, i.e.~$\rho_\mathrm{AB}^0 \not= \rho_\mathrm{A}^0 + \rho_\mathrm{B}^0$.

If the interaction between monomers is switched on we assume that the KS determinant of the dimer can be written as the antisymmetrized product of two determinants for both the monomers:
	\beq
		\tilde{\psi}_\mathrm{AB} = \asop \tilde{\psi}_\mathrm{A} \tilde{\psi}_\mathrm{B},
		\label{eq:PB_psiAB_inter}
	\eeq
and hence the dimer density fulfills the additivity condition~\eqref{eq:PB_trhoAB_add}. Owing to~\eqref{eq:PB_trhoAB_add} and using~\eqref{eq:PB_fundef_TsVneJExc}, the energy functional for the system corresponding to~\eqref{eq:PB_psiAB_inter} is
	\begin{align}
		& E_\mathrm{AB}[\rho_\mathrm{AB}] = E_\mathrm{AB}[\tilde{\rho}_\mathrm{AB}] = E_\mathrm{AB}[\tilde{\rho}_\mathrm{A} + \tilde{\rho}_\mathrm{B}] = \nonumber \\
		& = T^\mathrm{s}[\tilde{\rho}_\mathrm{A} + \tilde{\rho}_\mathrm{B}] + V_\mathrm{AB}^\mathrm{ne}[\tilde{\rho}_\mathrm{A} + \tilde{\rho}_\mathrm{B}] +
			J[\tilde{\rho}_\mathrm{A} + \tilde{\rho}_\mathrm{B}] + \nonumber \\
		& + E^\mathrm{xc}[\tilde{\rho}_\mathrm{A} + \tilde{\rho}_\mathrm{B}] + V^\mathrm{nn}_\mathrm{AB}. \label{eq:PB_EAB_rhoAB}
	\end{align}
Now we rewrite the functional~\eqref{eq:PB_EAB_rhoAB} extracting the monomer contributions to the dimer energy through a careful inspection of the terms in~\eqref{eq:PB_EAB_rhoAB}. It is clear from~\eqref{eq:PB_Ts_fun} that the non-interacting kinetic energy functional is linear,
	\beq
		T^\mathrm{s}[\tilde{\rho}_\mathrm{A} + \tilde{\rho}_\mathrm{B}] = T^\mathrm{s}[\tilde{\rho}_\mathrm{A}] + T^\mathrm{s}[\tilde{\rho}_\mathrm{B}],
	\eeq
the nuclear-electron attraction may be separated as
	\begin{align}
		& V_\mathrm{AB}^\mathrm{ne}[\tilde{\rho}_\mathrm{A} + \tilde{\rho}_\mathrm{B}] = \nonumber \\
		& = \int_{\mathbb{R}^3} \Big( v_\mathrm{A}^\mathrm{ne}(\mathbf{r}) + v_\mathrm{B}^\mathrm{ne}(\mathbf{r}) \Big) \Big( \tilde{\rho}_\mathrm{A}(\mathbf{r}) +
			\tilde{\rho}_\mathrm{B}(\mathbf{r}) \Big) \, d^3 \mathbf{r} = \nonumber \\
		& = V_\mathrm{A}^\mathrm{ne}[\tilde{\rho}_\mathrm{A}] + V_\mathrm{B}^\mathrm{ne}[\tilde{\rho}_\mathrm{B}] + \nonumber \\
		& + \int_{\mathbb{R}^3} v_\mathrm{B}^\mathrm{ne}(\mathbf{r}) \tilde{\rho}_\mathrm{A}(\mathbf{r}) \, d^3 \mathbf{r} +
			\int_{\mathbb{R}^3} v_\mathrm{A}^\mathrm{ne}(\mathbf{r}) \tilde{\rho}_\mathrm{B}(\mathbf{r}) \, d^3 \mathbf{r},
	\end{align}
and the coulombic term may be decomposed according to
	\begin{align}
		& J[\tilde{\rho}_\mathrm{A} + \tilde{\rho}_\mathrm{B}] = \nonumber \\
		& = \frac{1}{2} \int_{\mathbb{R}^3} \int_{\mathbb{R}^3} \Big( \tilde{\rho}_\mathrm{A}(\mathbf{r}_1) +
			\tilde{\rho}_\mathrm{B}(\mathbf{r}_1) \Big) \Big( \tilde{\rho}_\mathrm{A}(\mathbf{r}_2) + \tilde{\rho}_\mathrm{B}(\mathbf{r}_2) \Big) \times \nonumber \\
		& \times r_{12}^{-1} \, d^3 \mathbf{r}_1 d^3 \mathbf{r}_2 = \nonumber \\
		& = J[\tilde{\rho}_\mathrm{A}] + J[\tilde{\rho}_\mathrm{B}] +
			\int_{\mathbb{R}^3} \int_{\mathbb{R}^3} \frac{\tilde{\rho}_\mathrm{A}(\mathbf{r}_1) \tilde{\rho}_\mathrm{B}(\mathbf{r}_2)}{r_{12}} \, d^3 \mathbf{r}_1 d^3 \mathbf{r}_2.
	\end{align}
However, the explicit analytical form of~the xc~functional is unknown and its approximations depend on the functional used. Thus, we introduce the xc~energy non-additivity,~$\Delta E_\mathrm{xc}$:
	\beq
		\Delta E_\mathrm{xc}[\tilde{\rho}_\mathrm{A} + \tilde{\rho}_\mathrm{B}] =
			E^\mathrm{xc}[\tilde{\rho}_\mathrm{A} + \tilde{\rho}_\mathrm{B}] - E^\mathrm{xc}[\tilde{\rho}_\mathrm{A}] - E^\mathrm{xc}[\tilde{\rho}_\mathrm{B}].
		\label{eq:PB_Exc_nonadd}
	\eeq
It is worthwhile to note that the present formulation neither separates nor approximates any of the kinetic non-additivity terms appearing in the method of~\citet{wesol_warchel:1993}. These terms are implicitly and exactly included in the term~\eqref{eq:PB_Exc_nonadd} and thus are automatically accounted for in a consistent manner for any functional. Although the~expression~\eqref{eq:PB_EAB_rhoAB} is a functional of a~single density, we now make use of~\eqref{eq:PB_trhoAB_add} and treat the system energy as a bifunctional depending on both monomer densities:
	\beq
		E_\mathrm{AB}[\tilde{\rho}_\mathrm{A} + \tilde{\rho}_\mathrm{B}] \equiv E_\mathrm{AB}[\tilde{\rho}_\mathrm{A}; \tilde{\rho}_\mathrm{B}].
	\eeq
Thus, in our search for the ground-state dimer energy, we will minimize, with respect to~$\tilde{\rho}_\mathrm{A}$~and~$\tilde{\rho}_\mathrm{B}$, the bifunctional of the form:
	\begin{align}
		& E_\mathrm{AB}[\tilde{\rho}_\mathrm{A}; \tilde{\rho}_\mathrm{B}] = \nonumber \\
		& = T^\mathrm{s}[\tilde{\rho}_\mathrm{A}] + V_\mathrm{A}^\mathrm{ne}[\tilde{\rho}_\mathrm{A}] + J[\tilde{\rho}_\mathrm{A}] +
			E^\mathrm{xc}[\tilde{\rho}_\mathrm{A}] + V^\mathrm{nn}_\mathrm{A} + \nonumber \\
		& + T^\mathrm{s}[\tilde{\rho}_\mathrm{B}] + V_\mathrm{B}^\mathrm{ne}[\tilde{\rho}_\mathrm{B}] + J[\tilde{\rho}_\mathrm{B}] +
			E^\mathrm{xc}[\tilde{\rho}_\mathrm{B}] + V^\mathrm{nn}_\mathrm{B} + \nonumber \\
		& + \tilde{E}_\mathrm{int}[\tilde{\rho}_\mathrm{A}; \tilde{\rho}_\mathrm{B}] \nonumber \\
		& = E_\mathrm{A}[\tilde{\rho}_\mathrm{A}] + E_\mathrm{B}[\tilde{\rho}_\mathrm{B}] + \tilde{E}_\mathrm{int}[\tilde{\rho}_\mathrm{A}; \tilde{\rho}_\mathrm{B}],
		\label{eq:PB_EAB_bifun}
	\end{align}
where
	\begin{align}
		& \tilde{E}_\mathrm{int}[\tilde{\rho}_\mathrm{A}; \tilde{\rho}_\mathrm{B}] = \nonumber \\
		& = \int_{\mathbb{R}^3} v_\mathrm{B}^\mathrm{ne}(\mathbf{r}) \tilde{\rho}_\mathrm{A}(\mathbf{r}) \, d^3 \mathbf{r} +
			\int_{\mathbb{R}^3} v_\mathrm{A}^\mathrm{ne}(\mathbf{r}) \tilde{\rho}_\mathrm{B}(\mathbf{r}) \, d^3 \mathbf{r} + \nonumber \\
		& + \int_{\mathbb{R}^3} \int_{\mathbb{R}^3} \frac{\tilde{\rho}_\mathrm{A}(\mathbf{r}_1) \tilde{\rho}_\mathrm{B}(\mathbf{r}_2)}{r_{12}} \, d^3 \mathbf{r}_1 d^3 \mathbf{r}_2 +
			V_\mathrm{int}^\mathrm{nn} + \nonumber \\
		& + \Delta E_\mathrm{xc}[\tilde{\rho}_\mathrm{A}; \tilde{\rho}_\mathrm{B}] = \nonumber \\
		& = E_\mathrm{elst}[\tilde{\rho}_\mathrm{A}; \tilde{\rho}_\mathrm{B}] + \Delta E_\mathrm{xc}[\tilde{\rho}_\mathrm{A}; \tilde{\rho}_\mathrm{B}]. \label{eq:PB_tEint_bifun}
	\end{align}
In the above equation, $V_\mathrm{int}^\mathrm{nn}$ is intermonomer nuclear-nuclear repulsion energy. However, for the density additivity condition~\eqref{eq:PB_trhoAB_add} to hold, all orbitals must be kept mutually orthogonal, and the orthogonality also ensures that the intersystem Pauli exclusion principle is fulfilled. To this end, we perform the variational optimization in two steps, using the~Pauli blockade~(PB) method of~\citet{gut_pauli:1988}: first, the bifunctional extremal search is performed without the imposition of the intermonomer orthogonality constraint, and secondly, the penalty operator is added in the resulting iterative scheme. The penalty operator for monomer~A reads
	\beq
		\hat{\tilde{R}}_\mathrm{A} = \sum_{i \in A} \Ket{\tilde{a}_i} \Bra{\tilde{a}_i}
		\label{eq:PB_penfun}
	\eeq
and it is obvious that its action on monomer B's orbitals annihilates them once the orbitals are orthogonal. Now we turn to the first step: to find a~bifunctional minimum, we calculate the variational derivative of~\eqref{eq:PB_EAB_bifun} with respect to~$\tilde{\rho}_\mathrm{A}$:
	\begin{align}
		& \frac{\delta E_\mathrm{AB}[\tilde{\rho}_\mathrm{A}; \tilde{\rho}_\mathrm{B}]}{\delta \tilde{\rho}_\mathrm{A}(\mathbf{r})} = \nonumber \\
		& = \kinop + v_\mathrm{A}^\mathrm{ne}(\mathbf{r}) + \hat{\tilde{\jmath}}_\mathrm{A}(\mathbf{r}) + \tilde{v}_\mathrm{A}^\mathrm{xc}(\mathbf{r}) + \nonumber \\
		& + v_\mathrm{B}^\mathrm{ne}(\mathbf{r}) + \hat{\tilde{\jmath}}_\mathrm{B}(\mathbf{r}) + \Delta \tilde{v}_\mathrm{A}^\mathrm{xc}(\mathbf{r}) = \nonumber \\
		& = \hat{\tilde{f}}_\mathrm{A}^\mathrm{KS}(\mathbf{r}) + \Delta \tilde{v}_\mathrm{A}^\mathrm{xc}(\mathbf{r}) + \hat{\tilde{v}}_\mathrm{B}^\mathrm{elst}(\mathbf{r}),
	\end{align}
where the electrostatic potential is
	\beq
		\hat{\tilde{v}}_\mathrm{B}^\mathrm{elst}(\mathbf{r}) = v_\mathrm{B}^\mathrm{ne}(\mathbf{r}) + \hat{\tilde{\jmath}}_\mathrm{B}(\mathbf{r})
	\eeq
and the non-additivity xc~operator reads
	\begin{align}
		& \Delta \tilde{v}_\mathrm{A}^\mathrm{xc}(\mathbf{r}) = \nonumber \\
		& = \frac{\delta \Delta E^\mathrm{xc}[\tilde{\rho}_\mathrm{A}; \tilde{\rho}_\mathrm{B}]}{\delta \tilde{\rho}_\mathrm{A}(\mathbf{r})} = \nonumber \\
		& = \frac{\delta E^\mathrm{xc}[\tilde{\rho}_\mathrm{A} + \tilde{\rho}_\mathrm{B}]}{\delta \tilde{\rho}_\mathrm{A}(\mathbf{r})} -
			\frac{\delta E^\mathrm{xc}[\tilde{\rho}_\mathrm{A}]}{\delta \tilde{\rho}_\mathrm{A}(\mathbf{r})} = \nonumber \\
		& = \frac{\delta E^\mathrm{xc}[\rho_\mathrm{AB}]}{\delta \rho_\mathrm{AB}(\mathbf{r})} - \tilde{v}_\mathrm{A}^\mathrm{xc}(\mathbf{r}) = \nonumber \\
		& = v_\mathrm{AB}^\mathrm{xc}(\mathbf{r}) - \tilde{v}_\mathrm{A}^\mathrm{xc}(\mathbf{r}).
	\end{align}
Hence, the Euler equation for the bifunctional~\eqref{eq:PB_EAB_bifun} is
	\beq
		\mu_\mathrm{A} = \kinop + \hat{v}_\mathrm{A}^\mathrm{eff}(\mathbf{r})
	\eeq
with
	\beq
		\hat{v}_\mathrm{A}^\mathrm{eff}(\mathbf{r}) =
			\hat{\tilde{v}}_\mathrm{A}^\mathrm{elst}(\mathbf{r}) +
			\tilde{v}_\mathrm{A}^\mathrm{xc}(\mathbf{r}) +
			\hat{\tilde{v}}_\mathrm{B}^\mathrm{elst}(\mathbf{r}) +
			\tilde{v}_\mathrm{A}^\mathrm{xc}(\mathbf{r}) +
			\Delta \tilde{v}_\mathrm{A}^\mathrm{xc}(\mathbf{r}),
	\eeq
and~$\mu_\mathrm{A}$ being the Lagrange multiplier for the constraint:
	\beq
		N_\mathrm{A} - \int_{\mathbb{R}^3} \tilde{\rho}_\mathrm{A}(\mathbf{r}) \, d^3 \mathbf{r} = 0
	\eeq
The minimization of~\eqref{eq:PB_EAB_bifun} with respect to~$\tilde{\rho}_\mathrm{B}$ proceeds in an analogous way. Finally, the orbitals minimizing the functional~\eqref{eq:PB_EAB_bifun} are determined to satisfy
	\beq
		\begin{cases}
			\Big( \hat{\tilde{f}}_\mathrm{A}^\mathrm{KS}(\mathbf{r}) + \Delta \tilde{v}_\mathrm{A}^\mathrm{xc}(\mathbf{r}) + \hat{\tilde{v}}_\mathrm{B}^\mathrm{elst}(\mathbf{r}) \Big) \tilde{a}_i(\mathbf{r}) = \epsilon_{\mathrm{A}, i} \tilde{a}_i(\mathbf{r}) \\
			\Big( \hat{\tilde{f}}_\mathrm{B}^\mathrm{KS}(\mathbf{r}) + \Delta \tilde{v}_\mathrm{B}^\mathrm{xc}(\mathbf{r}) + \hat{\tilde{v}}_\mathrm{A}^\mathrm{elst}(\mathbf{r}) \Big) \tilde{b}_k(\mathbf{r}) = \epsilon_{\mathrm{B}, k} \tilde{b}_k(\mathbf{r})
		\end{cases}.
		\label{eq:PB_full_SC_ab}
	\eeq
In the second step of the PB procedure, the iterative process of solving Eqs.~\eqref{eq:PB_full_SC_ab} with the aid of the penalty operator is formulated. For monomer~A the $n$th iterative step reads
	\begin{align}
		& \Big( \hat{\tilde{f}}_\mathrm{A}^{\mathrm{KS}[n - 1]} + \Delta \tilde{v}_\mathrm{A}^{\mathrm{xc}[n - 1]} +
			\hat{\tilde{v}}_\mathrm{B}^{\mathrm{elst}[n - 1]} + \eta \hat{\tilde{R}}_\mathrm{B}^{[n - 1]} \Big) a_i^{[n]} = \nonumber \\
		& = \epsilon_{\mathrm{A}, i}^{[n]} a_i^{[n]}
		\label{eq:PB_full_SC_ab_iter}
	\end{align}
and its equivalent for monomer~B is obtained through the interchange of the A and~B subscripts in~\eqref{eq:PB_full_SC_ab_iter}. $\eta > 0$ is a parameter not affecting the final solutions. The orbitals obtained in~\eqref{eq:PB_full_SC_ab_iter} are orthogonalized, yielding an orthonormal $\left\{ \left\{ \tilde{a}_i^{[n]} \right\}_{i \in A}; \left\{ \tilde{b}_k^{[n]} \right\}_{k \in B} \right\}$ set. The interaction energy at the $n$th iteration is obtained upon the insertion of the densities calculated with these orbitals into~\eqref{eq:PB_EAB_bifun} and subtracting the unperturbed monomer energies:
	\begin{align}
		& E_\mathrm{int}^{\mathrm{PB}[n]} = E_\mathrm{AB} \left[ \tilde{\rho}_\mathrm{A}^{[n]}; \tilde{\rho}_\mathrm{B}^{[n]} \right]
			- E_\mathrm{A} \left[ \rho^0_\mathrm{A} \right] - E_\mathrm{B} \left[ \rho^0_\mathrm{B} \right] = \nonumber \\
		& = \Delta \tilde{E}_\mathrm{A}^{[n]} + \Delta \tilde{E}_\mathrm{B}^{[n]} +
			E_\mathrm{elst} \left[ \tilde{\rho}_\mathrm{A}^{[n]}; \tilde{\rho}_\mathrm{B}^{[n]} \right] +
			\Delta E_{\mathrm{xc}} \left[ \tilde{\rho}_\mathrm{A}^{[n]}; \tilde{\rho}_\mathrm{B}^{[n]} \right].
		\label{eq:PB_Eint_bifun}
	\end{align}
In the above equation, the A~monomer deformation is
	\beq
		\Delta \tilde{E}_\mathrm{A} = E_\mathrm{A} \left[ \tilde{\rho}_\mathrm{A}^{[n]} \right] - E_\mathrm{A} \left[ \rho^0_\mathrm{A} \right],
	\eeq
and analogously for monomer~B. Upon reaching self-consistency, the energy~\eqref{eq:PB_Eint_bifun} is equal to the supermolecular DFT interaction energy of~\eqref{eq:Eint_DFT}. The computational cost of our approach is essentially the same as that of the standard KS calculations.

Since the iterative process~\eqref{eq:PB_full_SC_ab_iter} starts with the~KS orbitals of the isolated monomers, the zero-iteration interaction energy may be viewed as an analog of the well-known HF-based Heitler-London interaction energy. Specifically, we define the DFT-based HL interaction energy as
	\beq
		E_\mathrm{int}^\mathrm{HL} = E_\mathrm{int}^{\mathrm{PB}[0]} =
			E_\mathrm{AB}\left[ \tilde{\rho}_\mathrm{A}^0; \tilde{\rho}_\mathrm{B}^0 \right] -
			E_\mathrm{A} \left[ \rho^0_\mathrm{A} \right] -
			E_\mathrm{B} \left[ \rho^0_\mathrm{B} \right].
		\label{eq:PB_Eint_HL}
	\eeq
This definition is equivalent to that proposed by~\citet{cyb_seversen:2003}. The difference between the self-consistent interaction energy and the HL interaction energy,
	\beq
		E_\mathrm{def}^\mathrm{PB} = E_\mathrm{int}^\mathrm{PB} - E_\mathrm{int}^\mathrm{HL},
		\label{eq:PB_Edef}
	\eeq
is referred to as the deformation energy.

It should be stressed here that both $E_\mathrm{int}^\mathrm{HL}$ and $E_\mathrm{def}^\mathrm{PB}$, as defined by the above equations, are uniquely defined, and are independent of the orthogonalization procedure, and may be interpreted in terms of SAPT. The $E_\mathrm{int}^\mathrm{HL}$ is the HL energy arising between the unperturbed, non-orthogonal monomers. At the HF level of theory, it includes the intermolecular electrostatic and exchange energies. In the DFT case, depending on a particular functional, it may also contain some obscure residual inter-monomer electron correlation terms that are related to the dispersion effect. This is because the interaction operator is, in general, the exchange-correlation operator rather than the exact-exchange one only, and the correlation is basically of a local type. 

The $E_\mathrm{def}^\mathrm{PB}$ term represents the deformation effect with respect to the non-orthogonal isolated monomers. First, it includes both the induction effects and the CT effects that are related to the induction and exchange-induction energies as defined by SAPT except that it is obtained iteratively through the infinite order rather than perturbatively through the finite order. Second, if an exact exchange-correlation operator were used the dispersion energy would be included in $E_\mathrm{def}^\mathrm{PB}$. As for now, the majority of existing functionals fail to to account for dispersion and at the same time they are not entirely dispersion-free. Consequently, for such functionals, $E_\mathrm{def}^\mathrm{PB}$ contains some residual dispersion terms as well.

\section{Numerical Results}

\subsection{Computational Details}

The method described in~Sec.~\ref{sec:theory} has been coded within the Molpro program suite~\cite{MOLPRO_brief}. Numerical calculations were carried out for three different functionals: Slater-Dirac~\cite{slater:1951}~(henceforth termed DIRAC), PBE0~\cite{adamo_barone:1999}, and B3LYP~\cite{becke:1993, steph_davlin_chab_frisch:1994}, and for model systems: two van der Waals complexes~(Ne$_2$ and the ethylene dimer), and two hydrogen-bonded dimers~[(HF)$_2$ and (NH$_3$)$_2$]. For comparison, we also present the results for the standard Hartree-Fock method~(HF). The distance between Ne~atoms in Ne$_2$ was set to~$6$~$a_0$. The geometries for (NH$_3$)$_2$ and (C$_2$H$_4$)$_2$ were taken from Ref.~\cite{jur_cerny_hobza:2006} and from Ref.~\cite{halk_klop_helg_jorg_taylor:1999} for~(HF)$_2$. The numerical procedure depends on the following parameters: energy threshold, i.e. the minimum difference in interaction energies from successive iterations for which the iterations are continued, the grid threshold, i.e. the accuracy with which the Slater-Dirac functional can be integrated using the grid as compared to its analytical integral, and $\eta$ parameter of~Eq.~\eqref{eq:PB_full_SC_ab_iter}. $\eta = 10^5$ was used for all calculations. However, we stress once more that $\eta$ does not affect the final solutions, only the convergence.

\begin{figure}[htbp]
	\bec
	\includegraphics[width = 8.6cm]{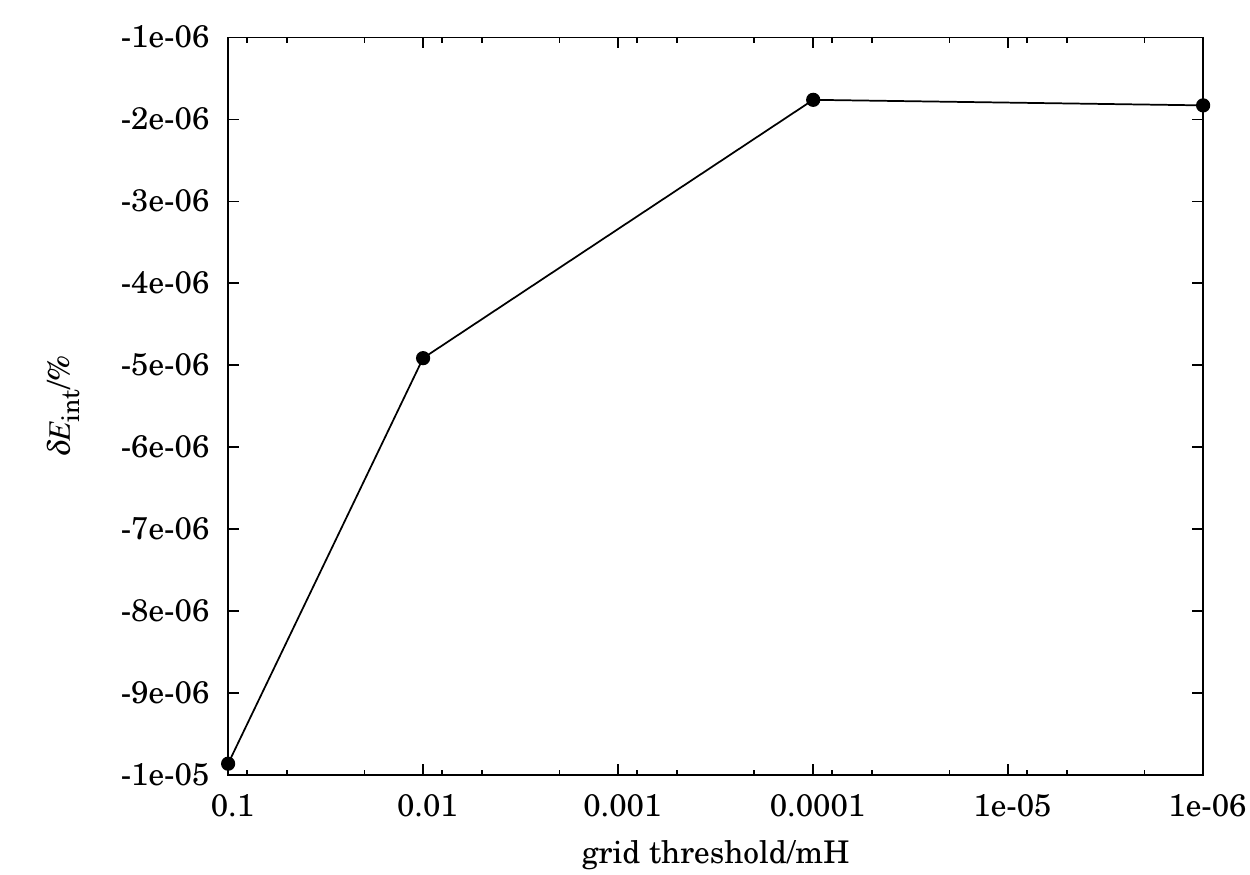}
	\caption{Dependence of $\delta E_\mathrm{int}$ on grid threshold for the Ne$_2$ dimer at $R = 6$~$a_0$ for PBE0 functional in aug-cc-pVQZ basis set.}
	\label{fig:thresh_deltas}
	\eec
	\end{figure}

In~Table~\ref{tab:en_bifun}
	\begin{table*}[htbp]
\caption{Interaction energies and their contributions for the bifunctional approach and its comparison with the DFT supermolecular energies. All values in~mH. The numbers in parentheses denote powers of ten.}
\label{tab:en_bifun}
\bec
\begin{tabular}{cc*{9}{>{$}c<{$}}}
\hline \hline
System & Functional & \Delta \tilde{E}_\mathrm{A} & \Delta \tilde{E}_\mathrm{B} & E_\mathrm{elst} & \Delta E_\mathrm{xc} & E_\mathrm{int}^\mathrm{HL} & E_\mathrm{def}^\mathrm{PB} & E_\mathrm{int}^\mathrm{PB} & \delta E_\mathrm{int}/\% \\
\hline
\multirow{4}{*}{\makecell{Ne$_2$ \\ $(R = 6$~$a_0)$}} & DIRAC & 0.658 & 0.658 & -0.953 & -0.625 & -0.262 & -0.0249 & -0.287 & -5.54(-5) \\

 & PBE0 & 0.416 & 0.416 & -0.587 & -0.343 & -0.098 & -0.00862 & -0.107 & -3.74(-6) \\

 & B3LYP & 0.463 & 0.463 & -0.664 & -0.155 & 0.108 & -0.00905 & 0.0993 & 2.8(-6) \\

 & HF & 0.238 & 0.238 & -0.325 & -0.0872 & 0.064 & -0.00148 & 0.0626 & 0.00014 \\

\hline
\multirow{4}{*}{(C$_2$H$_4$)$_2$} & DIRAC & 6.04  & 6.04  & -6.71 & -7.15 & -1.77 & -1.21 & -2.98 & -0.00285 \\

 & PBE0 & 4.58  & 4.58  & -5.11 & -3.98 & 0.0664 & -0.702 & -0.636 & -0.0412 \\

 & B3LYP & 4.86  & 4.86  & -5.37 & -2.82 & 1.54  & -0.708 & 0.834 & 0.047 \\

 & HF & 4.4   & 4.4   & -4.96 & -2.14 & 1.7   & -0.374 & 1.32  & 0.000301 \\

\hline
\multirow{4}{*}{(HF)$_2$} & DIRAC & 16.3  & 19.3  & -28.7 & -10.7 & -3.68 & -6.28 & -9.96 & -0.00402 \\

 & PBE0 & 13.3  & 15.9  & -24.7 & -6.91 & -2.31 & -4.98 & -7.28 & -0.00026 \\

 & B3LYP & 14.2  & 17    & -25.9 & -7    & -1.69 & -5.08 & -6.77 & -0.00223 \\

 & HF & 11    & 13.1  & -22.2 & -4.17 & -2.19 & -3.61 & -5.8  & -2.12(-5) \\

\hline
\multirow{4}{*}{(NH$_3$)$_2$} & DIRAC & 11.8  & 11.8  & -18.9 & -8.78 & -4.04 & -2.89 & -6.93 & -0.00268 \\

 & PBE0 & 9.4   & 9.4   & -15.7 & -5.49 & -2.36 & -2.06 & -4.41 & -0.000912 \\

 & B3LYP & 9.97  & 9.97  & -16.4 & -4.9  & -1.31 & -2.11 & -3.43 & -0.000516 \\

 & HF & 8.69  & 8.69  & -14.9 & -3.3  & -0.813 & -1.42 & -2.23 & -2.16(-6) \\

\hline \hline
\end{tabular}
\eec
\end{table*}
we present the numerical values of the components of~Eq.~\eqref{eq:PB_Eint_bifun} together with the DFT-based HL interaction energy~\eqref{eq:PB_Eint_HL}, the deformation~\eqref{eq:PB_Edef}, and the relative difference between the interaction energies obtained in~a bifunctional and supermolecular approaches,
	\beq
		\delta E_\mathrm{int} = \frac{E_\mathrm{int}^\mathrm{PB} - E_\mathrm{int}^\mathrm{DFT}}{E_\mathrm{int}^\mathrm{DFT}} \cdot 100\%.
		\label{eq:PB_dEint}
	\eeq
In Fig.~\ref{fig:thresh_deltas} we present the dependence of the relative difference~\eqref{eq:PB_dEint} on the grid threshold.
The energy threshold was set to~$10^{-9}$~mH and it was kept at that value for all values of the grid threshold reported in~Fig.~\ref{fig:thresh_deltas}. The calculations employed aug-cc-pVQZ basis set for neon dimer and aug-cc-pVTZ for the other systems~\cite{aug-cc-pVTQZ, feller:1996, sch_didier_els_sun_gur_chase_liand_windus:2007}. Dimer-centred basis set~(DCBS) has been used throughout the calculations and the supermolecular interaction energies were counterpoise~(CP)-corrected for the basis set superposition error~(BSSE).

\subsection{Discussion}

From the results of Table~\ref{tab:en_bifun} it is clear that the bifunctional approach converges to the same values of the interaction energies as the conventional KS procedure, within excellent accuracy of below 0.003~\% for the H-bonded dimers, and 0.05~\% for the van~der~Waals dimers.

In general, as long as our procedure is convergent, it must converge to the same result as the standard KS approach. This is because no extra constraints are imposed on the functional, and the finally optimized total density must be the same in both cases. However, the details of the convergence depend on several factors:
	\begin{itemize}
		\item the particular functional,
		\item the system under consideration,
		\item the initial orthogonalization of monomer orbitals, and
		\item the orthogonality forcing technique in the iteration process.
	\end{itemize}
The functional and system convergence dependence is evident in Table~\ref{tab:en_bifun}. The dependence on mutual orthognalization scheme and the manner it is forced in the iteration process have not been studied so far as only one approach has been adopted --- they will be studied in the future in the context of actual applications. For instance, convergence problems may appear for complexes which undergo major redistribution of electron densities between monomers, such as in donor-acceptor interactions and also for those which are poorly described by a single determinant due to static correlation effects.

The total interaction energy is composed of the HL energy and the PB deformation energy. As pointed out in the previous Section, both $E_\mathrm{int}^\mathrm{HL}$ and $E_\mathrm{def}^\mathrm{PB}$ are uniquely defined, and are independent of the orthogonalization procedure. However, this is not the case for the first four terms of~Eq.~\eqref{eq:PB_Eint_bifun} in~Table~\ref{tab:en_bifun}. They describe the monomer deformation effects due to orthogonalization (cols.~3 and~4), and electrostatic and exchange-correlation effects, arising between the monomers described with orthogonalised occupied orbitals (cols.~5 and~6). Therefore they are not uniquely defined, and are strongly dependent on the orthogonalization scheme --- they are not useful to interpret the interaction.

The total interaction energies listed in the Table~\ref{tab:en_bifun} are qualitatively correct only for polar molecules; for atoms and non-polar species they are erroneous due to the well known fact that the functionals: DIRAC, PBE0, B3LYP do not reproduce the dispersion component. The $E_\mathrm{int}^\mathrm{HL}$ and $E_\mathrm{def}^\mathrm{PB}$ components may be compared to similar terms at the HF level of theory (the last entry for every dimer). Assuming that the functionals reproduce only the local correlation terms, but fail to recover dispersion contributions, the DFT $E_\mathrm{int}^\mathrm{HL}$ results should differ from the HF ones by a relatively small intramonomer correlation effect. This is apparently not the case for DIRAC, for van der Waals complexes: Ne$_2$ and (C$_2$H$_4$)$_2$. Also PBE0 shows attraction, albeit small, for Ne$_2$, and seems to be not repulsive enough for (C$_2$H$_4$)$_2$. Such a behavior indicates that some residual dispersion terms are present. As to the B3LYP functional, it seems to be the most dispersion-free, since its values of $E_\mathrm{int}^\mathrm{HL}$ are the closest to the HF values. These results are in agreement with the observations that the amount of dispersion in a functional correlates to the steepness of the exchange enhancement factor: it becomes steeper when moving from DIRAC to B3LYP, and, consequently, the amount of dispersion included in these functionals is reduced~(see Refs.~\cite{lacks_gordon:1993, zhang_pan_yang:1997, wesol_parisel_ellinger_weber:1997, menconi_tozer:2005}) Finally, for hydrogen-bonded complexes, all methods give qualitatively correct $E_\mathrm{int}^\mathrm{HL}$, only PBE0 and B3LYP are closer to the HF result than DIRAC.

The above discussion suggests a useful application of the formalism presented in this paper. For approximate functionals the partitioning of the interaction energy into the $E_\mathrm{int}^\mathrm{HL}$ and $E_\mathrm{def}^\mathrm{PB}$ components may serve as a diagnostic of the functional's adequacy in the intermolecular interaction energy problems. Indeed, one can determine, what components and how efficiently are recovered by a tested functional.

\section{Summary and Outlook}

In this paper we provided a rigorous derivation of the supermolecular DFT interaction energy in terms of mutual monomer polarization via the Pauli blockade method. Numerical calculations for four model systems and three example functionals of different types have proved that the formalism leads to interaction energies rapidly converging to the supermolecular interaction energies. The accuracy achieved is better than 0.1~\% and appears to be limited only by the size of the grid. The accuracy is qualitatively similar for all three DFT functionals under investigation. Our formalism appears thus to be a viable and useful alternative of solving the KS equations in terms of separated-monomer orbitals rather than supermolecular orbitals. This fact has several important practical implications.

On the one hand, the presented formalism offers possibilities of using different functionals to describe the monomers and to describe the interaction. We have recently exploited this feature to design a novel DFT+dispersion approach~\cite{rajchel_zuch_szczesniak_chal:2009}. On the other, it would be of interest to combine our approach for the DFT that is capable to reproduce the dispersion energy, e.g. of range-separated family of functionals~\cite{angyan_gerber_savin_toul:2005}.

The bifunctional formulation provides a platform for deriving a choice of DFT treatments which use different functionals (or even theories) for different subsystems. The results may thus be of interest for those who use subsystem formulation in the context of embedding potentials~\cite{kev_dulak_wes:2006, cortona:1991, stef_truong:1996, kluner_govind_wang_carter:2001}.

Our formalism may also be useful when interpreting different interaction energy decomposition schemes that have recently been proposed within the DFT approach~\cite{cyb_seversen:2003, fren_wich_fr_los_lein_fr_ray:2003, reinh_piqueman_savin:2008, su_li:2009}.

\section{Acknowledgments}

Financial support from NSF (Grant~No.~CHE-0719260) is gratefully acknowledged. We acknowledge computational resources purchased through NSF~MRI program (Grant~No.~CHE-0722689). We are grateful to Bogumił Jeziorski and Maciej Gutowski for helpful discussion.


\bibliography{Paper-CPL-arXiv}

\end{document}